# ITIL frameworks to ITD Company for improving capabilities in service management

Phuc V. Nguyen
*Ho Chi Minh City's Vietnamese Communist Party Committee*
*x201102x@gmail.com*

**Abstract**

*IT operates in dynamic environments with the need always to change and adapt. There is a need to improve performance. Many gaps were found when we conduct the IT audit and we tried to seek to close gaps in capabilities. One way to the close these gaps is the adoption of good practices in wide industry use. There are several sources for good practices including public frameworks and standards such as ITIL, COBIT, CMMI, eSCM-SP, PRINCE2, ISO 9000, ISO/IEC 20000 and ISO/IEC 27001, etc. The paper propose ITIL frameworks to ITD Company for improving capabilities in service management.*

## 1. Introduction

Each solution has specific advantages and disadvantages. For the first and second solution, it requires more IT resource. Unfortunately, the company's business is focusing to trading, not software development. Moreover, the project team just support the company in a certain duration, not belong to the company. Thus, the two solutions are not suitable in this context. For the next solution, it requires a comprehensive change and huge investment, SAP ERP for instance. But they do not want to have a comprehensive change at one moment due to too many risks in change management and too much investment. Fortunately, the forth solution which can maximize usage of the existing systems and save cost have been chosen because it is aligned with the company situation. Therefore, they want to have a solution which can utilize the existing systems but it still enhance the current business.

## II. ITIL - Operations activities for IDT Company

In this portion, we will highlight some practices that will be proposed for ITIL implementation

- **Problem Management**: Perform root-cause analysis to resolve pervasive problems and prevent recurring incidents
- **Request Management**: Manage all user service requests to ensure effective service delivery
- **Change Management**: Execute approval workflows with full audit trail and secure e-mail approval with Change Management
- **Service Level Management**: Enforce and report on agreed upon response and resolution times based on your service level agreements and their related services
- **Contingency Management:** Manage issue and implement backup to prevent any failed in term of network, data, services,

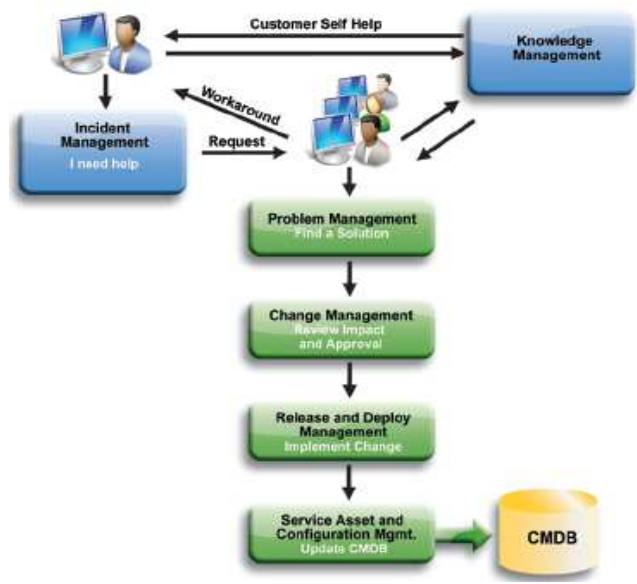

*Figure 1. Ideas for overcoming exiting gaps and improve system security*

## III. Resources owner & sharing

All system and resources should be controlled by an owner. This enables company to easy manage the system or resources effectively. So all change or resources modification must be approved before executing. In order to prevent data disclosure without control, the domain admin should be removed out of shared folder because IT staff can use it to read data. Here is the access permission table on the file server for each shared folder that need to implement

Other implement controls besides access permission above

- Domain Admin ID must be removed "Fully control" permission on shared folder, except "Backup Operators. This ID is used for backup activities
- Modify Domain policy to prevent "Backup Operators" log on to file server locally. This will prevent IT staff uses Backup operator to read information on file server.
- The password of backup operator ID should be kept by independent person, in that case HR manager or GD recommends keeping it.
- The password of domain admin ID also split into two parts and kept by two persons, it's only used if the change is approved and two persons will enter password seperately for IT staf further implement

## IV. Backup activity

All data that are being stored in the servers must be stored for backup. This prevents any data lost then may affect to business activities. The data is stored in the backup must be protected in high security and placed in safe box.

New backup driver and tapes should be purchased. The driver will be connected directly to files server. They are configured to copy whole data that stored in server (including EFFECT data, File server & Web data) into the tape daily based on defined schedule.

- HP/DELL backup driver DLT 40GB/80GB
- SCSI interface card
- Software: Windows backup or other applications
- Backup tape: Company needs to have at least two sets. There are 6 tapes in each set; they are labeled based on name of day in the week. All tapes have to be encrypted and the password for decrypting should be kept by independent person who are not relevant backup task. In case data restoring, the password keeper will enter password and IT staff will further process.
- Backup schedule: Incremental backup in daily tape & full backup on Saturday day. Backup sets are rotated in cycle. IT staff is responsible for daily replacing tape
- Backup tape storage: All tape must be stored in secure place such as cabinet with lock outside. The fully backup tape should be stored offsite company.

In the short term, while waiting for backup driver is purchased, IT should implement backup task by coping all data into another hard disk. To ensure all data at least stored separate two hard disk to prevent data lost due to hardware failed.

## V. Implement audit trail for EFFECT

All changes of master files and user permission access must be controlled by managers. In order to do that EFFECT system could be tracked all events and allows to printed out only by controller. The audit trail report is used to control the changes above.

By using SQL scripts, we will add it into each master database that need to be controlled. The script will be triggered to record all changed, including who change, when, what's changed, old value & new value of field, etc into new database named audit table. Not necessary to modify EFFECT application. The audit trail could be implemented from SQL server and report will be designed form crystal report.
Audit table and audit trail report is only used by auditor or controller, no one else could not access or modify.

So, in order to apply ITIL successfully, it requires many factors. Including top manager's support & involvement, well control, competent recourses, IT strategy, etc. We would start the ITIL framework at simplest level and then grow up step by step to adapt with company's appropriate situations.

## VI. The total security architecture

Security has applied for all levels as following:

- ❖ Protect from Internet: ISA server firewall, VPN and CISCO PIX 501
- ❖ Protection from virus: use Antivirus for clients and servers
- ❖ User access (authentication & authorization): use Active Directory, NTFS
- ❖ Database: use security of MS SQL Server 2005 and encrypted data (provided by SQL Server itself)

- Data loss: tape backup frequently
- Application: check username / password and access right

Hardware Architecture

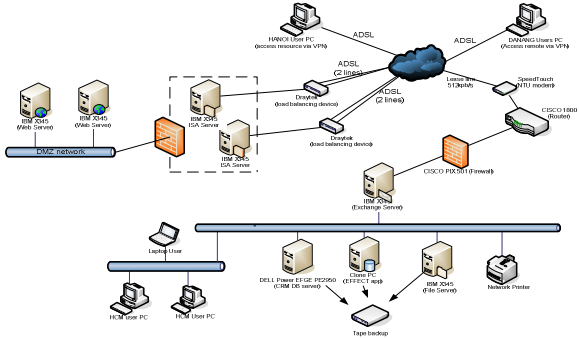

*Figure 2. The total hardware architecture*

Software basic Architecture

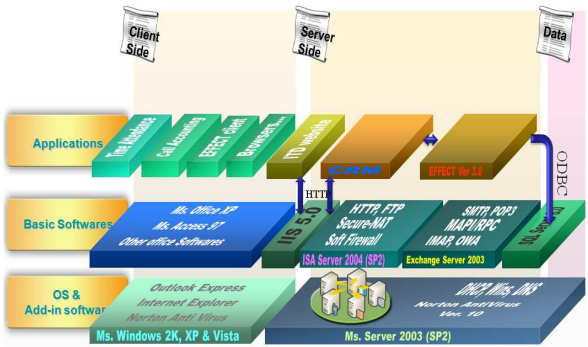

*Figure 3. The total software architecture*

Software Application Architecture

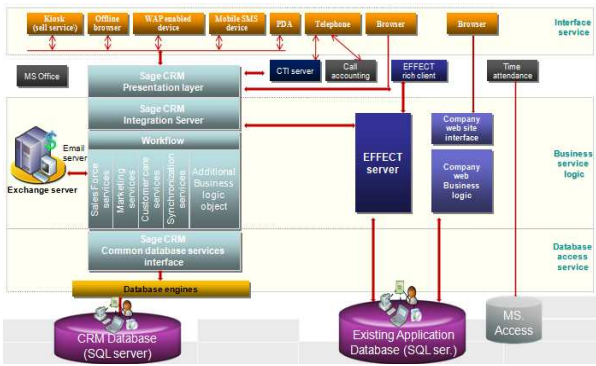

*Figure 4. The software application architecture*

## Conclusion

The above writing has presented approaches as well as aspects in project management such as technique requirement, objectives, scopes, WBS, schedule, workload and budget estimation, etc… The writing also mentioned to new solution for processes and direction of project development including short term, long term perspective.

In terms of short term, the company need to solve critical issues such as customer relationship enhancement and security. For those issues which need to be solved urgently, it requires our team conducting solutions. We have discussed various solutions like

- Develop a complete system
- Choose an open source and customize it
- Buy an available solution and apply it to the whole company's business
- Use a part of an available solution and integrate with existing systems